\documentclass[aps,pra,twocolumn,amsmath,amssymb,nofootinbib,superscriptaddress]{revtex4-1}
\usepackage{times}
\usepackage[pdftex]{graphicx}
\usepackage{dcolumn}
\usepackage{bm}
\usepackage{amsmath}
\usepackage{indentfirst}
\usepackage{float}
\usepackage[colorlinks]{hyperref}
\usepackage[dvipsnames]{xcolor}

\usepackage{makecell}

\newcommand{\gammaeff}{\gamma_{{\rm eff}}}
\newcommand{\Dop}{\mathcal{D}}
\newcommand{\Hop}{\hat{H}}
\newcommand{\rhoop}{\hat{\rho}}
\newcommand{\aop}{\hat{a}}
\newcommand{\adop}{\hat{a}^{\dagger}}
\newcommand{\xop}{\hat{x}}

\newcommand{\nave}{\bar{n}}

\newcommand{\trace}{{\rm tr}}
\newcommand{\im}{{\rm i}}

\begin{document}

\title{Fast Cooling of Trapped Ion in Strong Sideband Coupling Regime}

\author{Shuo Zhang}
\affiliation{Henan Key Laboratory of Quantum Information and Cryptography, Zhengzhou,
Henan 450000, China}

\author{Jian-Qi Zhang}
\affiliation{Innovation Academy for Precision Measurement Science and Technology, Wuhan, Hubei 430071, China}

\author{Wei Wu}
\affiliation{Department of Physics, College of Liberal Arts and Sciences, National University of Defense Technology, Changsha 410073, China
Interdisciplinary Center for Quantum Information, National University of Defense Technology, Changsha 410073, China}

\author{Wan-Su Bao}
\email{bws@qiclab.cn}
\affiliation{Henan Key Laboratory of Quantum Information and Cryptography, Zhengzhou,
Henan 450000, China}

\author{Chu Guo}
\email{guochu604b@gmail.com}
\affiliation{Henan Key Laboratory of Quantum Information and Cryptography, Zhengzhou,
Henan 450000, China}

\begin{abstract}
Trapped ion in the Lamb-Dicke regime with the Lamb-Dicke parameter
$\eta\ll1$ can be cooled down to its motional ground state using
sideband cooling. Standard sideband cooling works in the weak sideband
coupling limit, where the sideband coupling strength is small compared
to the natural linewidth $\gamma$ of the internal excited state,
with a cooling rate much less than $\gamma$. Here we consider 
cooling schemes in the strong sideband coupling regime, where the sideband
coupling strength is comparable or even greater than $\gamma$. We
derive analytic expressions for the cooling rate and the average occupation
of the motional steady state in this regime, based on which we show
that one can reach a cooling rate which is proportional to $\gamma$,
while at the same time the steady state occupation increases by a
correction term proportional to $\eta^{2}$ compared to the weak sideband
coupling limit. We demonstrate with numerical simulations that our
analytic expressions faithfully recover the exact dynamics in the
strong sideband coupling regime. 
\end{abstract}

\date{\today}
\maketitle

\address{}

\vspace{8mm}

\section{Introduction}

Trapped ions display rich physical phenomena due to its high degree
of controllability and abundant degrees of freedom. In the context
of quantum simulations, trapped ions can be used to realize quantum
spin systems~\cite{PorrasCirac2004b}, the Bose-Hubbard chain~\cite{PorrasCirac2004a},
the Jaynes-Cummings model~\cite{LeibfriedWineland2003} with tunable
interactions, as well as to study energy and particle transport far
from equilibrium~\cite{BermudezPlenio2013,RuizCampo2014,RammHartmut2014,GuoPoletti2015,GuoPoletti2016,GuoPoletti2017,GuoPoletti2017b,GuoPoletti2018}.
Trapped ions system is also a promising candidate to build quantum
computers~\cite{LanyonRoos2011,KielpinskiWineland2002}, where the
internal degrees of freedom are used to encode the qubits and the
external (motional) degrees of freedom are used to induce effective
couplings between those qubits~\cite{CiracZoller1995}. The motional
degrees of freedom of trapped ions play a central role in realizing
all the above systems or schemes, being used either directly or indirectly.
Particularly, in the context of quantum computing, cooling the motional
degrees of freedom down to their ground states is a central step for
coherent manipulation of the quantum state~\cite{WinelandMeekhof1998}.


Sideband cooling is one of the first and still widely used
cooling scheme~\cite{DiedrichWineland1989, MonroeWineland1995, RoosBlatt1999}. The key
ideas of sideband cooling are summarized as follows, which are also
helpful for understanding other cooling schemes using static lasers.
First we assume that an ion with two internal states, say a metastable
ground state $\vert g\rangle$ and an unstable excited state $\vert e\rangle$
with a lifetime $\tau=1/\gamma$ ($\gamma$ is the linewidth and we
have set $\hbar=1$), is trapped in a way that the motional degree
of freedom of the ion is described by a harmonic oscillator with equidistant
energy levels $\vert n\rangle$ of energies $(n+\frac{1}{2})\nu$,
where $\nu$ is the trap frequency. A laser with Rabi frequency
$\Omega$ is then applied onto the ion with detuning $\Delta$ which,
in the first order of the \textit{Lamb-Dicke} (LD) parameter $\eta$,
induces a \textit{carrier} transition $\vert g,n\rangle\leftrightarrow\vert e,n\rangle$
with strength $\Omega$, a \textit{red sideband} $\vert g,n\rangle\leftrightarrow\vert e,n-1\rangle$
and a \textit{blue sideband} $\vert g,n\rangle\leftrightarrow\vert e,n+1\rangle$
with strengths $\eta\Omega\sqrt{n}$ and $\eta\Omega\sqrt{n+1}$, respectively.
For this first order picture to be valid, the following conditions
need to be satisfied: 1) the Lamb-Dicke condition $\eta\ll1$, which
requires the oscillations of the trapped ion to be much smaller than
the wave length of the cooling laser, 2) resolved sideband condition,
which requires $\eta\Omega,\gamma\ll\nu$. The laser is often tuned
to \textit{red sideband resonance}, namely $\Delta=-\nu$, such that
the blue sideband is far-detuned compared to the red sideband and can often be neglected. 
The red sideband together with the natural decay $\vert e\rangle\rightarrow\vert g\rangle$
form a dissipative cascade~\cite{CohenTannoudji1998} which makes
the cooling possible. Most existing cooling schemes works in the weak
sideband coupling (WSC) regime, which requires additionally $\eta\Omega\ll\gamma$,
so that the states $\vert e,n-1\rangle$ decay to $\vert g,n-1\rangle$
immediately once pumped up from $\vert g,n\rangle$ by the red sideband.
As a result the states $\vert e,n-1\rangle$ can be adiabatically
eliminated from the cascade and one gets an effective decay from $\vert g,n\rangle$
to $\vert g,n-1\rangle$ with a rate 
\begin{align}
W_{{\rm WSC}} \propto \frac{\eta^{2}\Omega^{2}}{\gamma}.\label{eq:wsc}
\end{align}
The weak sideband coupling condition thus sets a cooling rate which
is much less than the natural linewidth $\gamma$. 

Subsequent proposals using static lasers mainly aim to improve the quality of
the steady state by suppressing the heating effects due to the carrier
transition or the blue sideband~\cite{CiracZoller1992,MorigiKeitel2000,Morigi2003,RetzkerPlenio2007},
or both of them~\cite{EversKeitel2004,CerrilloPlenio2010,CerrilloPlenio2018,AlbrechtPlenio2011,ZhangChen2012,ZhangChen2014,LuGuo2015},
by adding more lasers as well as more internal energy levels. 
As an outstanding example, cooling by electromagnetically induced
transparency (EIT) eliminates the carrier transition~\cite{MorigiKeitel2000,Morigi2003},
which has been demonstrated in various experiments due to its simplicity
and effectiveness~\cite{RoosBlatt2000,LinWineland2013,KampschulteMorigi2014,Lechner2016Roos,ScharnhorstSchmidt2018,JordanBollinger2019,FengMonroe2020,QiaoKim2020}. Quantum control has
also been applied in recent years to numerically find an optimal sequence
of pulsed lasers which drives the trapped ion towards its motional ground
state~\cite{MachnesRetzker2010}. An advantage of cooling with pulsed lasers
is that the lasers could often be tuned such that fast cooling can be achieved compared to  sideband cooling, while
the drawbacks are that the time-dependence of the lasers adds more
difficulty for the experimental implementation, and that the initial
motional state is often required to be known precisely in such schemes.


In this work, we focus on cooling schemes using static lasers. In
particular, we aim to improve the cooling rate set by Eq.(\ref{eq:wsc}),
which is essential in all applications to reduce the associated dead
time~\cite{ScharnhorstSchmidt2018}. For this goal, we consider
cooling schemes in the strong sideband coupling (SSC) regime, where $\eta\Omega$
is comparable with or even larger than $\gamma$, namely $\eta\Omega\geq\gamma$.
Eq.(\ref{eq:wsc}) fails in the SSC regime, which effect has been
observed both numerically~\cite{AlbrechtPlenio2011,ZhangChen2014,LuGuo2015,CerrilloPlenio2018}
and experimentally~\cite{ScharnhorstSchmidt2018}. In particular,
we consider both the standing wave sideband cooling and the EIT cooling
schemes in the SSC regime such that the heating effects due to the
carrier transitions are suppressed. In both cases, we show analytically
and numerically that we can reach a cooling rate $W_{\textrm{SSC}}\propto\gamma$
in the SSC regime, independent of the sideband coupling strength $\eta\Omega$.
The price to pay is a correction term to the steady state occupation
of the motional degree of freedom which is proportional to $\eta^{2}$.
This paper is organized as follows. We first consider the standing
wave sideband cooling in the SSC regime in Sec.~\ref{sec:swcooling}.
We derive analytic expressions for the dynamics of the average
occupation of the motional state as well as its steady state value,
which are verified by comparison to exact numerical results. In Sec.\ref{sec:eitcooling},
we further generalize those expressions to EIT cooling in the SSC
regime. We conclude in Sec.~\ref{sec:summary}.

\section{Standing wave sideband cooling in the SSC regime} \label{sec:swcooling} 

\begin{figure}
\begin{centering}
\includegraphics[width=1\columnwidth]{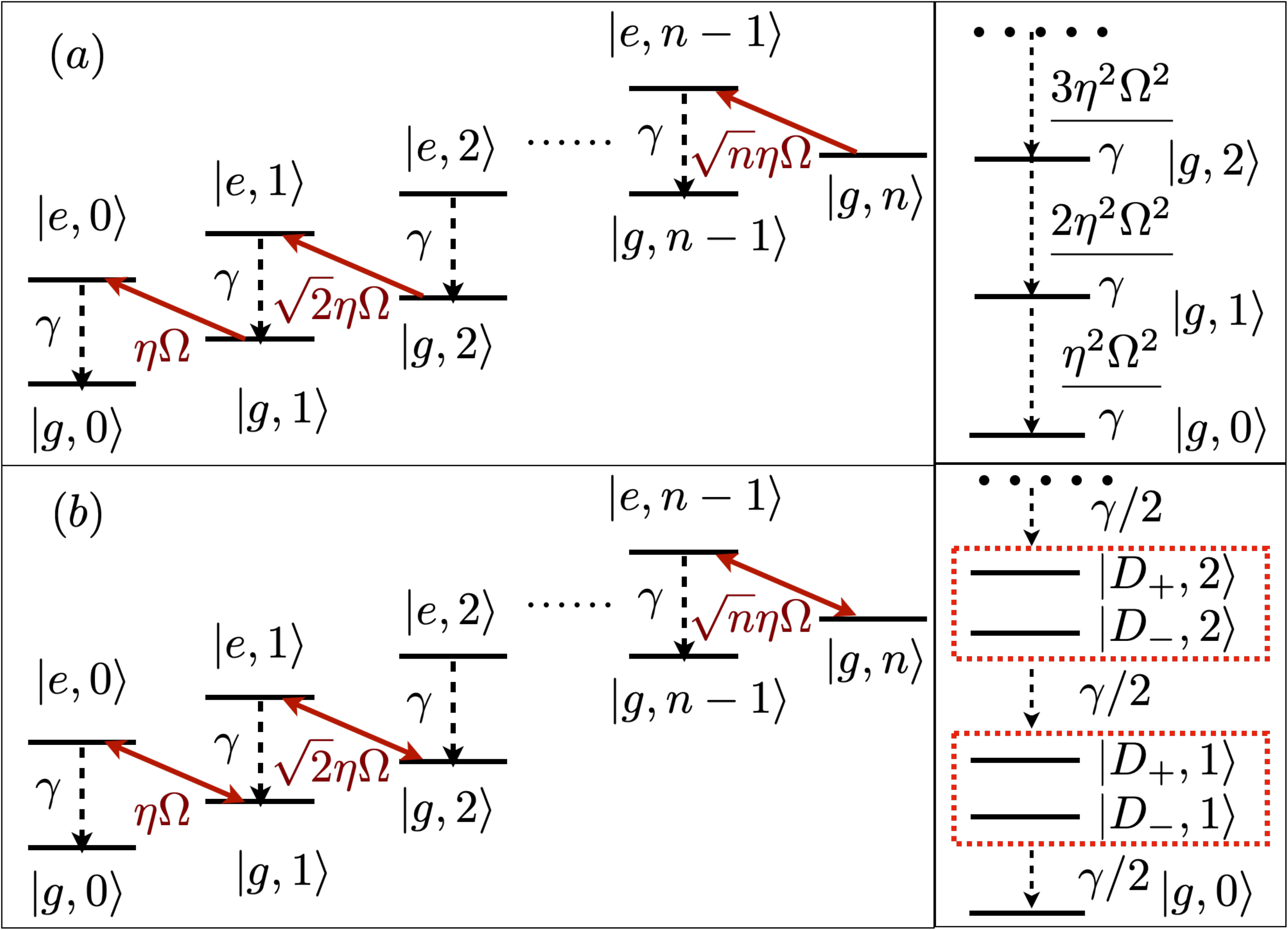}
\par\end{centering}
\caption{(a) Standing wave sideband cooling in the weak sideband coupling regime.
The laser resonantly couples $\vert g,n\rangle$ to $\vert e,n-1\rangle$,
which then decays to $\vert g,n-1\rangle$ immediately without oscillating
back into $\vert g,n\rangle$. Therefore the time to stay in the state
$\vert e,n-1\rangle$ is negligible and $\vert e,n-1\rangle$ is adiabatically
eliminated, resulting in an effective decay from $\vert g,n\rangle$
to $\vert g,n-1\rangle$ as shown in the box on right-hand side, with
a rate in Eq.(\ref{eq:wsc}). (b) Standing wave sideband cooling in
the strong sideband coupling regime. The occupation oscillates between
$\vert g,n\rangle$ and $\vert e,n-1\rangle$ before decaying into
$\vert g,n-1\rangle$. The cooling dynamics can be understood using
the two dressed states $\vert D_{+},n\rangle$ and $\vert D_{-},n\rangle$
defined in Eqs.(\ref{eq:dressedstates}), which forms a sub-block as
shown in the box on the right hand side. Each sub-block decays to a
lower block with an effective rate $\gamma/2$. }
\label{fig:fig1}
\end{figure}

Standing wave sideband cooling is conceptually
the simplest cooling scheme where the carrier transition is suppressed~\cite{CiracZoller1992}.
Here we first consider this scheme both due to its simplicity and that it is also helpful for understanding other cooling schemes that based on a dark state.
In standing wave sideband cooling scheme a trapped ion with a mass $m$ and two internal states,
a metastable ground state $\vert g\rangle$ with energy $\omega_{g}$
and an unstable excited state $\vert e\rangle$ with linewidth $\gamma$
and energy $\omega_{e}$, is coupled to a standing wave laser with
frequency $\omega_{L}$, wave number $k$ and Rabi frequency $\Omega$.
The ion is assumed to be trapped in a harmonic potential with a
trap frequency $\nu$ and at the same time located at the node of
the standing wave laser such that the carrier transition vanishes.
The equation of motion is described by the Lindblad master equation~\cite{GoriniSudarshan1976,Lindblad1976}
\begin{align}
\frac{\textrm{d}}{\textrm{d}t}\rhoop=-\im\left[\Hop_{\textrm{SW}},\rhoop\right]+\Dop_{\textrm{SW}}(\rhoop),\label{eq:swmaster}
\end{align}
where the Hamiltonian $\Hop_{\textrm{SW}}$ takes the form
\begin{align}\label{eq:swh}
\Hop_{\textrm{SW}}= & -\Delta\vert e\rangle\langle e\vert+\nu\adop\aop\nonumber \\
 & +\frac{\Omega}{2}(\vert g\rangle\langle e\vert+\vert e\rangle\langle g\vert)\sin(k\xop).
\end{align}
Here $\Delta=\omega_{L}-\left(\omega_{e}-\omega_{g}\right)$ is the
detuning, $\adop$ and $\aop$ are the creation and annihilation operators
for the motional state (the phonons), $\xop=\frac{1}{\sqrt{2m\nu}}(\adop+\aop)$
is the position operator. The dissipation takes the form
\begin{align}\label{eq:swd}
\Dop_{\textrm{SW}}(\rhoop)= & \gamma\int_{-1}^{1}d\left(\cos(\theta)\right)\left(\frac{3}{4}\left(1+\cos^{2}(\theta)\right)\right)\vert g\rangle\langle e\vert\nonumber \\
 & e^{\im k\xop\cos(\theta)}\rhoop e^{-\im k\xop\cos(\theta)}\vert e\rangle\langle g\vert-\frac{\gamma}{2}\{\vert e\rangle\langle e\vert,\rhoop\},
\end{align}
In the LD regime, the Hamiltonian $\Hop_{\textrm{SW}}$ can be approximated
as
\begin{align}
\Hop_{\textrm{SW}}^{{\rm LD}}= & -\Delta\left|e\right\rangle \left\langle e\right|+\nu\adop\aop\nonumber \\
 & +\frac{\eta\Omega}{2}\left(\left|g\right\rangle \left\langle e\right|+\left|e\right\rangle \left\langle g\right|\right)\left(\aop+\adop\right)
\end{align}
by expanding $\sin(k\xop)$ to the first order of the LD parameter $\eta=k/\sqrt{2m\nu}$ in Eq.(\ref{eq:swh}).
The dissipator $\Dop_{\textrm{SW}}$ is often kept to the zeroth order
of $\eta$, which is
\begin{align}
\Dop_{\textrm{SW}}^{{\rm LD}}(\rhoop)=\gamma\left(\left|g\right\rangle \left\langle e\right|\rhoop\left|e\right\rangle \left\langle g\right|-\frac{1}{2}\{\rhoop,\left|e\right\rangle \left\langle e\right|\}\right),\label{eq:swdld}
\end{align}
since the higher orders terms only contributes in order $\eta^{4}$~\cite{AlbrechtPlenio2011}.
When the standing wave laser is tuned to red sideband resonance, namely
$\Delta=-\nu$, we further neglect the blue sideband as a first approximation since it
is far-detuned compared to the red sideband in the resolved sideband
regime. Thus we
are left with the following approximate Hamiltonian in the interacting
picture
\begin{align}
\Hop_{\textrm{SW}}^{{\rm LDR}}=\frac{\eta\Omega}{2}\left(\vert g\rangle\langle e\vert\adop+\vert e\rangle\langle g\vert\aop\right).\label{eq:swhldr}
\end{align}
As a result, Eq.(\ref{eq:swmaster}) is approximated by
\begin{align}
\frac{\textrm{d}}{\textrm{d}t}\rhoop=-\im\left[\Hop_{\textrm{SW}}^{{\rm LDR}},\rhoop\right]+\Dop_{\textrm{SW}}^{{\rm LD}}(\rhoop).\label{eq:swmasterld}
\end{align}
In the WSC regime, namely when $\eta\Omega\ll\gamma$, the time required
by the transition between the state $\vert g,n\rangle$ and the state
$\vert e,n-1\rangle$ is much larger than the life time of $\vert e,n-1\rangle$,
which means that once the state $\vert e,n-1\rangle$ is populated
by the red sideband, it immediately decays to the state $\vert g,n-1\rangle$
without being pumped back into $\vert g,n\rangle$, as shown in the
left hand side box of Fig.~\ref{fig:fig1}(a). The state $\vert e,n-1\rangle$
can thus be adiabatically eliminated and one obtains an effective
decay from $\vert g,n\rangle$ to $\vert g,n-1\rangle$ with an effective
decay rate as in Eq.(\ref{eq:wsc}), which is also shown in the right hand side
box of Fig.~\ref{fig:fig1}(a). 
In contrast, in the SSC regime, the state $\vert e,n-1\rangle$ may
oscillate back into the state $\vert g,n\rangle$ before decay into
$\vert g,n-1\rangle$ as shown in the left hand side box of Fig.~\ref{fig:fig1}(b).
Therefore it is more convenient to work in the dressed state representation with
\begin{subequations}\label{eq:dressedstates}
\begin{align}
\vert D_{+},n\rangle & =\frac{\sqrt{2}}{2}\left(\vert g,n\rangle+\vert e,n-1\rangle\right);\\
\vert D_{-},n\rangle & =\frac{\sqrt{2}}{2}\left(\vert g,n\rangle-\vert e,n-1\rangle\right),
\end{align}
\end{subequations} such that the two states $\vert D_{+},n\rangle$
and $\vert D_{-},n\rangle$ are eigenstates of $\Hop_{\textrm{SW}}^{{\rm LDR}}$.
To solve Eq.(\ref{eq:swmasterld}) in the SSC regime, we further assume
that the quantum state $\rhoop$ can be approximated by the following
ansatz
\begin{align}\label{eq:ansatz}
\rhoop(t)=a_{0}\left(t\right)\left|g,0\right\rangle \left\langle g,0\right|+\sum_{n=1}^{\infty}\left(b_{+,n}\left(t\right)\left|D_{+},n\right\rangle \left\langle D_{+},n\right|\right.\nonumber \\
\left.+b_{-,n}\left(t\right)\left|D_{-},n\right\rangle \left\langle D_{-},n\right|\right),
\end{align}
that is, only the diagonal terms in the dressed state representation
is considered. To this end, we note that the elimination of the carrier transition is important since otherwise $\vert g, n\rangle$ will form a dressed state with $\vert e,n\rangle$ instead of $\vert e,n-1\rangle$ due to the fact that the carrier transition is much stronger than the red sideband, in which case our ansatz in Eq.(\ref{eq:ansatz}) will no longer be valid. Substituting Eq.(\ref{eq:ansatz}) into Eq.(\ref{eq:swmasterld}),
we get the equations for $b_{+,n}$, $b_{-,n}$ and $a_0$ 
\begin{subequations}
\begin{align}
\frac{\textrm{d}}{\textrm{d}t}b_{+,n} & =-\frac{\gamma}{2}b_{+,n}+\frac{\gamma}{4}\left(b_{+,n+1}+b_{-,n+1}\right);\\
\frac{\textrm{d}}{\textrm{d}t}b_{-,n} & =-\frac{\gamma}{2}b_{-,n}+\frac{\gamma}{4}\left(b_{+,n+1}+b_{-,n+1}\right);\\
\frac{\textrm{d}}{\textrm{d}t}a_{0} & =\frac{\gamma}{2}\left(b_{+,1}+b_{-,1}\right).
\end{align}
\end{subequations} Now we define $p_{n}=b_{+,n}+b_{-,n}$ and get
the equation for $p_{n}$ \begin{subequations} \label{eq:evolvep}
\begin{align}
\frac{\textrm{d}}{\textrm{d}t}p_{0} & =\frac{\gamma}{2}p_{1};\\
\frac{\textrm{d}}{\textrm{d}t}p_{n} & =\frac{\gamma}{2}p_{n+1}-\frac{\gamma}{2}p_{n}\quad\left(n>0\right).
\end{align}
\end{subequations} 
The equation of motion for the average phonon
number, defined as $\nave(t)=\trace(n\rhoop(t))=\sum_{n=1}^{\infty}\left(n-\frac{1}{2}\right)p_{n}(t)$, is then
\begin{align}
\frac{\textrm{d}}{\textrm{d}t}\nave(t)=-\frac{\gamma}{2}\left(1-p_{0}\left(t\right)\right)+\frac{1}{2}\frac{\textrm{d}}{\textrm{d}t}p_{0}\left(t\right).
\end{align}

In the following we assume that the initial state of the trapped ion is
\begin{align}
\rhoop_{0}=\left|g\right\rangle \left\langle g\right|\otimes\rhoop_{\textrm{th}}^{e},
\end{align}
where $\rhoop_{\textrm{th}}^{e}=\sum_{n=0}^{\infty}c_{n}\left|n\right\rangle \left\langle n\right|$
is a thermal state for the motional degree of freedom with average
phonon number $n_{0}$, that is, $c_{n}=n_{0}^{n}/\left(1+n_{0}\right)^{n+1}$~\cite{roosthesis}.
Due to the strong sideband coupling, the state $\vert g,n\rangle$
will be rapidly mixed with the state $\vert e,n-1\rangle$ at the
beginning of the cooling dynamics. As a result, after this very short
initial dynamics, the system will look as if it starts from another
initial state
\begin{align}
\rhoop_{0}'= & c_{0}\left(0\right)\left|g,0\right\rangle \left\langle g,0\right|\nonumber \\
 & +\sum_{n=1}^{\infty}\frac{c_{n}}{2}\left(\left|D_{+},n\right\rangle \left\langle D_{+},n\right|+\left|D_{-},n\right\rangle \left\langle D_{-},n\right|\right).
\end{align}
Compared to our ansatz in Eq.(\ref{eq:ansatz}), we can see that $a_{0}(0)=c_{0}$,
$b_{+,n}(0)=b_{-,n}(0)=c_{n}/2$. Then we can solve Eqs.(\ref{eq:evolvep})
with the initial conditions $p_{0}(0)=c_{0}$, $p_{n}(0)=c_{n}$,
and get
\begin{align}
p_{0}\left(t\right) & =1-\frac{n_{0}}{1+n_{0}}e^{-\frac{\gamma}{2}\frac{1}{1+n_{0}}t}.
\end{align}
Then we have
\begin{align}
\nave(t)=n_{0}'e^{-\frac{\gamma}{2}\frac{1}{1+n_{0}}t}\label{eq:nave}
\end{align}
with $n_{0}'=\left(n_{0}-\frac{n_{0}}{2(1+n_{0})}\right)$. We can
identify from Eq.(\ref{eq:nave}) that the cooling rate in the SSC
regime is
\begin{align}
W_{{\rm SSC}}^{{\rm SW}}=\frac{\gamma}{2}\frac{1}{n_{0}+1}.\label{eq:ssc}
\end{align}
There are several important differences between Eq.(\ref{eq:ssc})
derived in the SSC limit and Eq.(\ref{eq:wsc}) derived in the WSC
limit. First, $W_{{\rm SSC}}^{{\rm SW}}$ is proportional to the natural
linewidth $\gamma$ and is independent of the sideband coupling $\eta\Omega$.
This is because the effect of $\eta\Omega$ has already been absorbed
into the ansatz in Eq.(\ref{eq:ansatz}), where $\vert g,n\rangle$
is fully mixed with $\vert e,n-1\rangle$. Moreover, the dressed states $\vert D_{\pm,n}\rangle$
decay to $\vert D_{\pm,n-1}\rangle$ with an effective decay rate
of $\gamma/4$. As a result each $p_{n}$ (with $n\geq1$) decays
with a rate proportional to $\gamma/2$, as shown in the right hand
side box of Fig.~\ref{fig:fig1}(b). 
Second, $W_{{\rm SSC}}^{{\rm SW}}$ is inversely proportional to the
initial average photon number $n_{0}$, while in the WSC limit the
cooling rate is independent of $n_{0}$.

The steady state occupation predicted by Eq.(\ref{eq:nave}) is $\nave_{{\rm st}}=\nave(\infty)=0$,
this is because we have neglected all the heating terms in Eq.(\ref{eq:swmaster}).
In fact, when $\nave(t)$ approaches $0$, the blue sideband can no
longer be neglected since there is no red sideband for the state $\vert g,0\rangle$.
To reasonably evaluate $\nave_{{\rm st}}$, we first assume that the
trapped ion has already been cooled close to the ground state, namely
$\nave(t)\approx0$, such that we can limit ourself to the subspace
spanned by $\left\{ \left|g,0\right\rangle ,\left|g,1\right\rangle ,\left|e,0\right\rangle ,\left|e,1\right\rangle \right\} $.
Then we can employ the $4-$level Bloch equation for this subspace, which is,
\begin{subequations}\label{eq:bloch}
\begin{align}
\frac{d}{dt}\rho_{g0,g0} & =  \frac{\eta\Omega}{2}\sigma_{g0,e1}^{y}+\gamma\rho_{e0,e0}\\
\frac{d}{dt}\rho_{e0,e0} & =  \frac{\eta\Omega}{2}\sigma_{e0,g1}^{y}-\gamma\rho_{e0,e0}\\
\frac{d}{dt}\sigma_{g0,e1}^{y} & =  -2\nu\sigma_{g0,e1}^{x}-\eta\Omega\rho_{g0,g0}+\eta\Omega\rho_{e1,e1}-\frac{\gamma}{2}\sigma_{g0,e1}^{y}\\
\frac{d}{dt}\sigma_{g0,e1}^{x} & =  2\nu\sigma_{g0,e1}^{y}-\frac{\gamma}{2}\sigma_{g0,e1}^{x}\\
\frac{d}{dt}\rho_{e1,e1} & =  -\frac{\eta\Omega}{2}\sigma_{g0,e1}^{y}-\gamma\rho_{e1,e1}\\
\frac{d}{dt}\sigma_{e0,g1}^{y} & =  \eta\Omega\left(\rho_{g1,g1}-\rho_{e0,e0}\right)-\frac{\gamma}{2}\sigma_{e0,g1}^{y}.
\end{align}
\end{subequations}
Here we have used $\rho_{a,a}=\textrm{Tr}\left(\rhoop\left|a\right\rangle \left\langle a\right|\right)$,
$\sigma_{a,b}^{y}=\textrm{Tr}\left(\rhoop\left(i\left|a\right\rangle \left\langle b\right|-i\left|b\right\rangle \left\langle a\right|\right)\right)$,
$\sigma_{a,b}^{x}=\textrm{Tr}\left(\rhoop\left(\left|a\right\rangle \left\langle b\right|+\left|b\right\rangle \left\langle a\right|\right)\right)$, with $g0, g1, e0,e1$ 
standing for the states $\left|g,0\right\rangle$, $\left|g,1\right\rangle$, $\left|e,0\right\rangle$, $\left|e,1\right\rangle $ respectively. By solving Eqs.(\ref{eq:bloch}), we get the steady state populations for $\rho_{e0,e0}$, $\rho_{g1,g1}$, $\rho_{e1,e1}$ as
\begin{subequations}
\begin{align}
\rho_{e0,e0} & =  \frac{\left(\eta\Omega\right)^{2}}{\left(\eta\Omega\right)^{2}+16\nu^{2}+\gamma^{2}}\rho_{g0,g0}\\
\rho_{g1,g1} & =  \frac{\left(\eta\Omega\right)^{2}+\gamma^{2}}{\left(\eta\Omega\right)^{2}+16\nu^{2}+\gamma^{2}}\rho_{g0,g0}\\
\rho_{e1,e1} & =  \frac{\left(\eta\Omega\right)^{2}}{\left(\eta\Omega\right)^{2}+16\nu^{2}+\gamma^{2}}\rho_{g0,g0}.
\end{align}
\end{subequations}
Since the system has already been cooled close to its motional ground state,
i.e, $\rho_{g0,g0}\approx1$, and under the condition that $\nu\gg\eta\Omega,\gamma$,
the steady state phonon occupation is
\begin{align}\label{eq:nss}
\nave_{\textrm{st,SSC}}^{{\rm SW}} & =  \rho_{g1,g1}+\rho_{e1,e1} =  \frac{1}{8}\left(\frac{\eta\Omega}{\nu}\right)^{2}+\frac{\gamma^{2}}{16\nu^{2}}.
\end{align}
The last term in Eq.(\ref{eq:nss}), $\frac{\gamma^{2}}{16\nu^{2}}$, is exactly the steady state occupation of standing wave sideband cooling in the WSC regime, namely $\nave_{\textrm{st},\textrm{WSC}}^{{\rm SW}}=\frac{\gamma^{2}}{16\nu^{2}}$. The $\eta^2$ correction term also persists for $\nave_{\textrm{st},\textrm{WSC}}^{{\rm SW}}$ but is often neglected since in the WSC regime it is much smaller compared to $\frac{\gamma^{2}}{16\nu^{2}}$.
Eq.(\ref{eq:nave}) can thus be corrected as
\begin{align}\label{eq:navecorr}
\nave(t)=\left(n_{0}'-\nave_{\textrm{st, SSC}}^{{\rm SW}}\right)e^{-\frac{\gamma}{2}\frac{1}{1+n_{0}}t}+\nave_{\textrm{st,SSC}}^{{\rm SW}}.
\end{align}

\begin{figure}
\noindent \begin{centering}
\includegraphics[width=1\columnwidth]{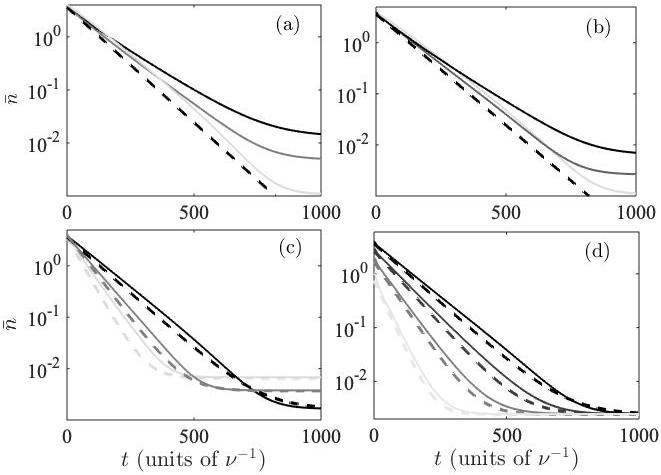}
\par\end{centering}
\caption{Average phonon occupation $\nave$ for the standing wave sideband
cooling as a function of time $t$. In all the panels $\Delta=-\nu$.
(a) The black lines from darker to lighter stand for the exact dynamics
at $\eta=0.2,0.12,0.04$, while the black dashed line is our analytic
prediction. The other parameters used are $n_{0}=4$, $\Omega=1.5\nu$,
$\gamma=0.1\nu$. (b) The black lines from darker to lighter stand
for the exact dynamics at $\Omega=2.1\nu,1.25\nu,0.6\nu$ respectively,
while the black dashed line is our analytic prediction. The other
parameters used are $n_{0}=4$, $\eta=0.1$, $\gamma=0.1\nu$. (c)
The black lines from darker to lighter stand for the exact dynamics at $\gamma=0.1\nu,0.15\nu,0.2\nu$ respectively with $\Omega=9\gamma$, while the dashed lines from
darker to lighter are the corresponding analytic predictions. The other parameters used are
$n_{0}=4$, $\eta=0.1$. (d) The black lines from darker to lighter
stand for the exact dynamics at $n_{0}=4,3,2,1$ respectively, while the dashed lines from
darker to lighter are the corresponding analytic predictions.
The other parameters used are $\eta=0.08$, $\Omega=1.5\nu$, $\gamma=0.1\nu$. }
\label{fig:fig2}
\end{figure}

To verify our physical picture in the SSC regime, we compare our analytic
expression in Eq.(\ref{eq:navecorr}) with the numerical solutions
of the exact Lindblad equation as in Eq.(\ref{eq:swmaster}). Concretely,
We plot the dependence of $\nave$ as a function of time $t$ in Fig.~\ref{fig:fig2}
for different values of $\eta$ (panel a), $\Omega$ (panel b), $\gamma$
(panel c) and $n_{0}$ (panel d) respectively. From Fig.~\ref{fig:fig2}(a,
b) we can see that our analytic expression works better for smaller
values of $\eta\Omega/\nu$, which is as expected since Eq.(\ref{eq:navecorr})
is derived based on the resolved sideband condition. In Fig.~\ref{fig:fig2}(c),
we have chosen different values of $\gamma$ such that $\eta\Omega/\nu=0.9\gamma/\nu\ll1$
is satisfied, and we can see that the dynamics predicted by Eq.(\ref{eq:navecorr})
agrees well with the exact numerical solutions, except that Eq.(\ref{eq:navecorr})
predicts a slightly faster decay since heating due to the blue sideband
is neglected in Eq.(\ref{eq:swmasterld}). In Fig.\ref{fig:fig2}(d),
we fix $\eta\Omega/\nu=1.2\gamma/\nu=0.12$ and we can see that the
cooling rate indeed has a strong dependence on $\nave_{0}$ as predicted
by Eq.(\ref{eq:nss}). The numerical simulations throughout this work
are done using the open source numerical package QuTiP~\cite{qutip}.


\begin{figure}
\noindent \begin{centering}
\includegraphics[width=1\columnwidth]{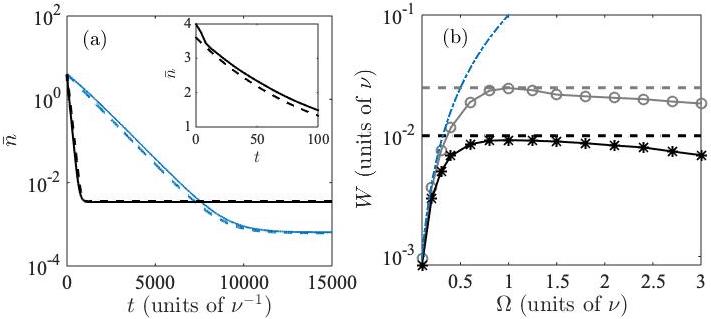}
\par\end{centering}
\caption{Average phonon occupation $\bar{n}$ as a function of time $t$ for
the standing wave sideband cooling. The black line and black dashed
line are exact and analytic results in the strong sideband coupling
limit with $\Omega=1.5\nu$. The blue line and blue dashed line are
exact and analytic results in the weak sideband coupling limit with
$\Omega=0.1\nu$. The inset highlights the short time dynamics in the strong sideband coupling limit. We have used $n_{0}=4$ in this panel.
(b) The darker black line with star and the darker black dashed line
show the exact and analytic cooling rate $W$ as a function of $\Omega$
with $n_{0}=4$, where the exact cooling rate is computed from the
exponential fitting of the exact dynamics. The lighter black line
with circle and the the lighter black dashed line show the exact and
analytic cooling rate $W$ as a function of $\Omega$ with $n_{0}=1$.
The blue dot-dashed line is the cooling rate predicted in the weak sideband
coupling limit by Eq.(\ref{eq:wsc}), which is independent of $n_{0}$.
The other parameters used in both panels are $\eta=0.1$, $\Delta=-\nu$,
$\gamma=0.1\nu$.}
\label{fig:fig3}
\end{figure}

To highlight the sharp differences between the SSC limit and the WSC
limit, we compare the dynamics as well as the steady state phonon occupation
in both regimes. In Fig.~\ref{fig:fig3}(a), we plot $\nave$ as
a function of $t$ in the SSC limit (the black line with $\eta\Omega=1.5\gamma$)
and in the WSC limit (the blue line with $\eta\Omega=0.1\gamma$),
the black and blue dashed lines are the corresponding analytic predictions.
In particular, we can see that the cooling rate in the SSC limit is
$10$ times larger than in the WSC limit, while the steady state phonon occupation
is $5.5$ times larger. In the inset of Fig.~\ref{fig:fig3}(a), we plot the short time dynamics in the SSC regime, from which we can see that at the beginning of the cooling dynamics, there is indeed a sudden drop of the average phonon occupation due to the formation of dressed states by rapidly mixing $\vert g,n\rangle$ with $\vert e,n-1\rangle$. In Fig.~\ref{fig:fig3}(b)
we plot the cooling rate, which results from an exponential fitting of
the exact dynamics, as a function of the Rabi frequency $\Omega$.
The darker black line with star corresponds to $n_{0}=4$ while the
lighter black line with circle corresponds to $n_{0}=1$. The darker
and lighter black dashed lines are the corresponding analytic predictions
from Eq.(\ref{eq:ssc}). The blue dot-dashed line is the prediction from
Eq.(\ref{eq:wsc}). We can see that Eq.(\ref{eq:wsc}) agrees well
with the numerical fitting when $\Omega/\nu<0.5$, where the WSC condition
is satisfied, and then for $1<\Omega/\nu<1.5$, the analytic predictions
from Eq.(\ref{eq:ssc}) agree well with the numerical fitting. For
even larger $\Omega$ such that $\Omega/\nu>2$, the derivation between
Eq.(\ref{eq:ssc}) and the numerical fitting becomes larger since
the resolved sideband condition is no longer satisfied.

\begin{figure}
\noindent \begin{centering}
\includegraphics[width=0.9\columnwidth]{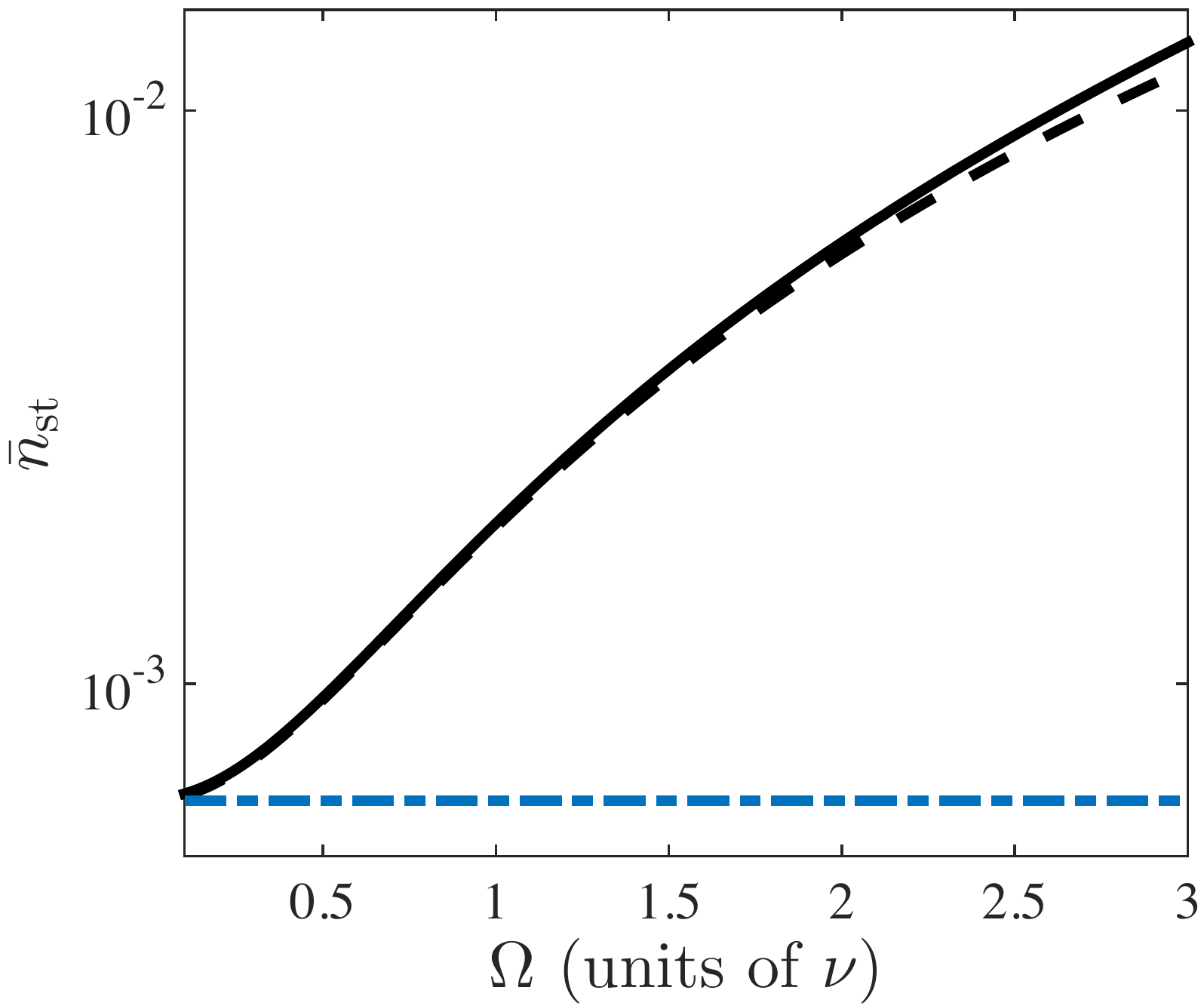}
\par\end{centering}
\caption{The steady state phonon occupation $\nave_{{\rm st}}$ as a function of the Rabi frequency $\Omega$. The black line and black dashed lines are exact numerical results and our analytic predictions in the strong sideband coupling regime, while the blue dot-dashed represents the analytic results predicted in the weak sideband coupling regime. The other parameters used are $n_0=4$, $\eta=0.1$, $\gamma=0.1\nu$, $\Delta=-\nu$. }
\label{fig:figst}
\end{figure}

Finally in Fig.~\ref{fig:figst}, we plot the steady state phonon occupation $\nave_{{\rm st}}$ as a function of $\Omega$. Interestingly, we can see that our analytic prediction in Eq.(\ref{eq:nss}) agrees well with the exact numerical results in all regimes (the derivations which happen for large $\Omega/\nu$ is because that the resolved sideband condition is no longer satisfied.). This is because that to derive Eq.(\ref{eq:nss}) we have kept both the red sideband and the blue sideband terms, and only assumed that the final occupation is close to $0$. The large derivation of $\nave_{\textrm{st},\textrm{WSC}}^{{\rm SW}}$ (the blue dot-dashed line) from the exact result (the black line) again signifies that the $\eta^2$ correction term can no longer be neglected when the WSC condition is not satisfied.

\section{EIT cooling in the SSC limit} \label{sec:eitcooling}

Standing wave sideband cooling, being conceptually simple, may have
several drawbacks in experimental implementations: 1) The standing
wave laser may not be easy to implement and 2) the natural linewidth
$\gamma$ is not a tunable parameter and thus the resolved sideband
condition $\gamma\ll\nu$ may not be satisfied. The EIT cooling
scheme overcomes both difficulties, while at the same time eliminates
the carrier transition.

\begin{figure}
\noindent \begin{centering}
\includegraphics[width=1\columnwidth]{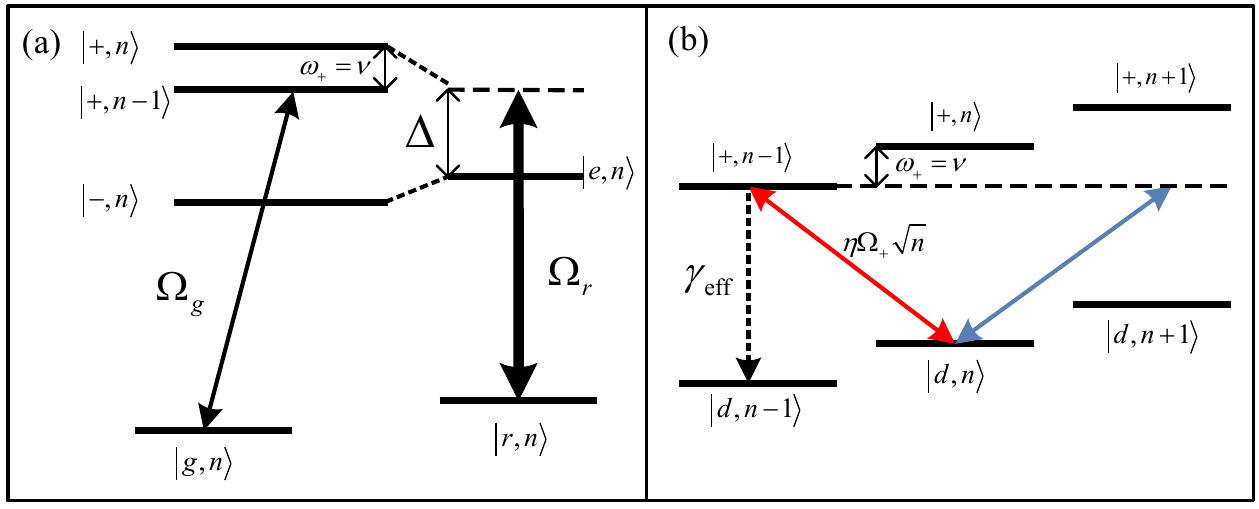}
\par\end{centering}
\caption{(a) The level diagram of EIT cooling. Two lasers couples $\vert g\rangle$
and $\vert r\rangle$ with strengths $\Omega_{g}$ and $\Omega_{r}$
to $\vert e\rangle$, with the same detuning $\Delta$. The internal
Hamiltonian containing the three levels $\vert g\rangle$, $\vert r\rangle$
and $\vert e\rangle$ is diagonalized and results in three dressed
state $\vert d\rangle$, $\vert+\rangle$ and $\vert-\rangle$ with
energies $\omega_{d}$, $\omega_{+}$, $\omega_{-}$, from lower to
higher, as in Eqs.(\ref{eq:eitbasis}, \ref{eq:eitenergies}). The
red sideband resonance condition is satisfied in the dressed state
picture by choosing $\omega_{+}=\nu$. (b) After neglecting the far-off
resonant coupling from $\vert d,n\rangle$ to $\vert-,n\rangle$,
the EIT cooling forms a dissipative cascade similar to the standing
wave sideband cooling. Strong sideband coupling condition is satisfied
if $\eta\Omega\gg\gammaeff$. }
\label{fig:fig4}
\end{figure}

A standard EIT cooling scheme uses a $\Lambda-$type three-level internal
structure with an excited state $\left|e\right\rangle $ of linewidth
$\gamma$, and two metastable ground states $\vert g\rangle$ and
$\vert r\rangle$. Two lasers are used, which induce transitions $\vert g\rangle\leftrightarrow\vert e\rangle$
and $\vert r\rangle\leftrightarrow\vert e\rangle$, with frequencies
$\omega_{g}$ and $\omega_{r}$, wave numbers $k_{g}$ and $k_{r}$,
Rabi frequencies $\Omega_{g}$ and $\Omega_{r}$ respectively. The internal level structure of EIT cooling is shown in Fig.~\ref{fig:fig4}(a). The
Hamiltonian of EIT cooling can thus be written as
\begin{align}
\Hop_{{\rm EIT}}= & -\Delta\vert e\rangle\langle e\vert+\nu\adop\aop\nonumber \\
 & +\frac{\Omega_{g}}{2}\left(\vert g\rangle\langle e\vert e^{-\im k_{g}\xop}+\vert e\rangle\langle g\vert e^{\im k_{g}\xop}\right)\nonumber \\
 & +\frac{\Omega_{r}}{2}\left(\vert r\rangle\langle e\vert e^{-\im k_{r}\xop}+\vert e\rangle\langle r\vert e^{\im k_{r}\xop}\right),
\end{align}
with $\eta_{g}=k_{g}/\sqrt{2m\nu}$ and $\eta_{r}=k_{r}/\sqrt{2m\nu}$,
and $\Delta$ being the detuning for both lasers. Similar to Eq.(\ref{eq:swd}),
the dissipative part of the EIT cooling can be written as 
\begin{align}
 & \Dop_{\textrm{EIT}}(\rhoop)\nonumber \\
= & \sum_{j=g,r}\gamma_{j}\int_{-1}^{1}d\left(\cos(\theta)\right)\left(\frac{3}{4}\left(1+\cos^{2}(\theta)\right)\right)\vert j\rangle\langle e\vert\nonumber \\
 & e^{\im k_{j}\xop\cos(\theta)}\rhoop e^{-\im k_{j}\xop\cos(\theta)}\vert e\rangle\langle j\vert-\frac{\gamma_{g}}{2}\{\vert e\rangle\langle e\vert,\rhoop\}.
\end{align}
where $\gamma_{g}$ and $\gamma_{r}$ are the decay rates from $\vert e\rangle$
to $\vert g\rangle$ and $\vert r\rangle$ respectively.

The internal degrees of freedom of $\Hop_{{\rm EIT}}$ can be diagonalized
with three eigenstates~\cite{FleischhauerJonathan2005}
\begin{subequations}\label{eq:eitbasis}
\begin{align}
\left|+\right\rangle  & =\sin\phi\left|e\right\rangle -\cos\phi\left(\sin\vartheta\left|g\right\rangle +\cos\vartheta\left|r\right\rangle \right);\\
\left|-\right\rangle  & =\cos\phi\left|e\right\rangle +\sin\phi\left(\sin\vartheta\left|g\right\rangle +\cos\vartheta\left|r\right\rangle \right);\\
\left|d\right\rangle  & =\cos\vartheta\left|g\right\rangle -\sin\vartheta\left|r\right\rangle
\end{align}
\end{subequations} with energies \begin{subequations}\label{eq:eitenergies}
\begin{align}
\omega_{+} & =\frac{1}{2}\left(-\Delta+\sqrt{\Omega_{r}^{2}+\Omega_{g}^{2}+\Delta^{2}}\right);\\
\omega_{-} & =\frac{1}{2}\left(-\Delta-\sqrt{\Omega_{r}^{2}+\Omega_{g}^{2}+\Delta^{2}}\right);\\
\omega_{d} & =0.
\end{align}
\end{subequations} Here, the angles $\phi$ and $\vartheta$ are
defined by 
\begin{align}
\tan2\phi & =-\frac{\sqrt{\Omega_{r}^{2}+\Omega_{g}^{2}}}{\Delta};\\
\tan\vartheta & =\frac{\Omega_{g}}{\Omega_{r}}.
\end{align}
The EIT cooling condition is chosen as $\omega_{+}=\nu$ such that
effective red sideband $\left|d,n\right\rangle \rightarrow\left|+,n-1\right\rangle $
is resonant. In the dressed state basis as defined in Eqs.(\ref{eq:eitbasis}),
and neglecting the far-detuned state $\left|-\right\rangle $
as well as the effective blue sideband $\vert d,n\rangle\leftrightarrow\vert+,n+1\rangle$,
we get an effective Hamiltonian in the interacting picture 
\begin{align}\label{eq:eithld}
\Hop_{{\rm EIT}}^{{\rm LDR}}= i\eta_{D}\frac{\Omega_{+}}{2}\left(\left|d\right\rangle \left\langle +\right|\adop-\left|+\right\rangle \left\langle d\right|\aop\right),
\end{align}
with $\eta_{D}=\eta_{g}-\eta_{r}$, and $\Omega_{+}=\frac{\Omega_{g}\Omega_{r}}{\sqrt{\Omega_{r}^{2}+\Omega_{g}^{2}}}\sin\phi$.
The effective dissipation in the dressed state basis is 
\begin{align}
\Dop_{{\rm EIT}}^{{\rm LD}}(\rhoop)=\gamma_{\textrm{eff}}\left(\left|d\right\rangle \left\langle +\right|\rhoop\left|+\right\rangle \left\langle d\right|-\frac{1}{2}\{\rhoop,\left|+\right\rangle \left\langle +\right|\}\right)\label{eq:eitdld}
\end{align}
with
\begin{align}
\gamma_{\textrm{eff}}=\frac{\sin^{2}\phi}{2}\left(\gamma_{g}\cos^{2}\vartheta+\gamma_{r}\sin^{2}\vartheta\right).
\end{align}
Comparing Eqs.(\ref{eq:eithld}, \ref{eq:eitdld}) with Eqs.(\ref{eq:swhldr},
\ref{eq:swdld}), we can see that the EIT cooling is equivalent to
the standing wave sideband cooling, by making the substitutions $\eta\rightarrow\eta_{D}$, $\Omega\rightarrow\Omega_{+}$
$\gamma\rightarrow\gammaeff$, which is shown in Fig.~\ref{fig:fig4}(b).
As a result, the cooling rate and the
steady state phonon occupation for EIT cooling in the SSC regime $\eta_{D}\Omega_{+}\geq\gamma_{\textrm{eff}}$ can be read from Eqs.(\ref{eq:ssc}, \ref{eq:nss}) as
\begin{align}
W_{\textrm{SSC}}^{{\rm EIT}} & =\frac{\gamma_{\textrm{eff}}}{2}\frac{1}{1+n_{0}};\label{eq:eitscc}\\
\nave_{\textrm{st,SSC}}^{{\rm EIT}} & =\frac{1}{8}\left(\frac{\eta_{D}\Omega_{\text{+}}}{\nu}\right)^{2}+\nave_{\textrm{st,WSC}}^{{\rm EIT}}. \label{eq:eitnss}
\end{align}
with $\nave_{\textrm{st,WSC}}^{{\rm EIT}}=\left(\frac{\gamma}{4\Delta}\right)^{2}$
being steady state average phonon occupation of EIT cooling in WSC
regime.

\begin{figure}
\noindent \begin{centering}
\includegraphics[width=1\columnwidth]{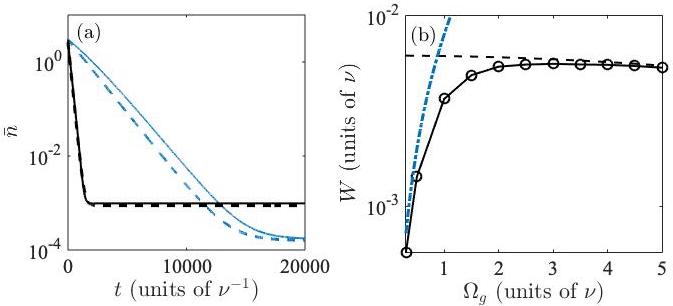}
\par\end{centering}
\caption{(a) The average photon number $\nave$ as a function of $t$ for EIT
cooling. The black and black dashed lines are exact and analytic results
in the strong sideband coupling limit with $\Omega_{g}=4\nu$. The
blue and blue dashed lines are exact and analytic results in the weak
sideband coupling limit with $\Omega_{g}=0.3\nu$. (b) Cooling rate
$W$ as a function of the Rabi frequency $\Omega$. The black dashed
line represents the analytic predictions from Eq.(\ref{eq:eitsscapp}) and the black line with circuit
represent the cooling rate from an exponential fitting of the exact dynamics. The blue dashed line is the
cooling rate in the weak sideband coupling limit as in Eq.(\ref{eq:eitwsc}). The other parameters used in both panels are $n_{0}=3$,
$\eta_{g}=0.1$, $\eta_{r}=-0.1$, $\Delta=103\nu$, $\gamma_{g}=5\nu$,
$\gamma_{r}=0$, $\Omega_{r}=20\nu$.}
\label{fig:fig5}
\end{figure}

In the experimental implementations of EIT cooling, the coupling strengths are 
usually chosen such that $\Omega_{g}\ll\Omega_{r}$ \cite{MorigiKeitel2000,RoosBlatt2000}.
As a result the internal dark state $\vert d\rangle\approx\left|g\right\rangle $.
Therefore, we have $\gamma_{\textrm{eff}}\approx\frac{\nu}{\Delta}\gamma_{g}$
and $\Omega_{+}\approx\frac{\Omega_{g}\Omega_{r}}{2\Delta}$, and
the cooling rate in SSC regime becomes
\begin{align}
W_{\textrm{SSC}}^{{\rm EIT}}\approx\frac{\gamma_{g}\nu}{2\Delta}\frac{1}{1+n_{0}}.\label{eq:eitsscapp}
\end{align}
Similar to Eq.(\ref{eq:ssc}), we can see that the cooling rate is
mainly determined by $\gamma_{g}$. In contrast, in the WSC regime,
the cooling rate is related to $\gamma=\gamma_{g}+\gamma_{r}$ as
\begin{align}
W_{\textrm{WSC}}^{{\rm EIT}} \approx \eta_{D}^{2}\frac{\Omega_{g}^{2}}{\gamma},\label{eq:eitwsc}
\end{align}
where in the condition $\Omega_{g}\ll\Omega_{r}$ has been used in deriving
the above equation.

Similar to Fig.~\ref{fig:fig3}, we compare the sharp difference
between EIT cooling in the SSC limit and in the the WSC limit in Fig.~\ref{fig:fig5}.
In Fig.~\ref{fig:fig5}(a), we plot the $\nave$ as a function of
time in the SSC limit (the black line with $\eta_{D}\Omega_{+}\approx3.3\gammaeff$)
and in the WSC limit (the blue line with $\eta_{D}\Omega_{+}=0.25\gammaeff$),
the black and blue dashed lines are the corresponding analytic predictions, where we can see that our analytic expressions in Eqs.(\ref{eq:eitscc}, \ref{eq:eitnss}) agree very well with the exact numerical results, and that the cooling rate in the SSC regime is indeed much faster than that in the WSC regime.
In Fig.~\ref{fig:fig5}(b) we plot the cooling rate resulting from
an exponential fitting of the exact dynamics as a function of the
Rabi frequency $\Omega_{g}$. We note that in such parameters settings
we have $\Omega_{g}\approx\frac{2\Omega_{+}\Delta}{\Omega_{r}}\approx10\Omega_{+}$
and $\gammaeff\approx0.05\nu$. The black line with circle and the
black dashed line correspond to the exact numerical results and the analytic predictions
from Eq.(\ref{eq:eitsscapp}) respectively, with $n_{0}=3$. The
blue dot-dashed line stands for the analytical predictions from Eq.(\ref{eq:eitwsc}). We
can see that Eq.(\ref{eq:eitwsc}) agrees well with the exact numerical results
when $\Omega_{g}/\nu<0.5$ (corresponding to $\eta_{D}\Omega_{+}/\gammaeff<0.2$ where the WSC condition is satisfied). While for $\Omega_{g}/\nu>3$
(corresponding to $\eta_{D}\Omega_{+}/\gammaeff>1.2$ where the SSC condition is satisfied), our analytic
predictions from Eq.(\ref{eq:eitsscapp}) agree well with the exact numerical results.

\section{Conclusion} \label{sec:summary} 

In summary, we have studied standing wave sideband
cooling and EIT cooling of trapped ion in the strong sideband coupling
regime. We derived analytic expressions for the cooling dynamics as
well as for the steady state occupation of the motional state in the strong sideband coupling regime, showing
that in this regime we could reach a cooling
rate which is proportional to the linewidth $\gamma$ of the excited
state ,
and which also depends on the initial occupation $n_{0}$ of the motional
state. 
This is in comparison with current weak sideband coupling based
cooling schemes where the cooling rate is much smaller than $\gamma$
and is independent of $n_{0}$. Additionally, the steady state occupation of the motional state
increases by a term proportional to $\eta^{2}$ compared to the weak
sideband coupling limit. The analytic expressions are verified against
the numerical results by solving the exact Lindblad master equation,
showing that they could faithfully recover both the short time and
long time dynamics for the motional state of the trapped ion. Our results
could be experimentally implemented to speed up the cooling of a trapped ion by a factor
of $10$ compared to current weak sideband coupling based schemes
such as EIT cooling, and can be easily extended to other dark-state based cooling schemes.

\section{Acknowledgement}
S. Z acknowledges support from National Natural Science Foundation of China under Grant No. 11504430. C. G acknowledges support from National Natural Science Foundation of China under Grant No. 11805279.


%

\end{document}